\documentclass[preprint]{elsarticle}

\usepackage{epsfig}
\usepackage{amssymb}
\usepackage{midfloat}
\usepackage{bm}
\usepackage[dvips]{color}

\begin{document}

\begin{frontmatter}

\title{Measurement of the neutron electric dipole moment via spin
  rotation in a non-centrosymmetric crystal}

\author[PNPI]{V.V.~Fedorov}
\author[ILL]{M.~Jentschel}
\author[PNPI]{I.A.~Kuznetsov}
\author[PNPI]{E.G.~Lapin}
\author[ILL]{E.~Leli{\`e}vre-Berna}
\author[ILL]{V.~Nesvizhevsky}
\author[ILL]{A.~Petoukhov}
\author[PNPI]{S.Yu.~Semenikhin}
\author[ILL]{T.~Soldner\corref{cor2}}
\author[PNPI]{V.V.~Voronin\corref{cor1}}
\author[PNPI]{Yu.P. Braginetz}

\address[PNPI]{Petersburg Nuclear Physics Institute, Gatchina,
  St.Petersburg, Russia}
\address[ILL]{Institut Laue-Langevin, Grenoble, France}

\cortext[cor2]{soldner@ill.fr. Present address: Physik-Department E18,
  Technische Universit{\"a}t M{\"u}nchen, Garching, Germany.}
\cortext[cor1]{vvv@pnpi.spb.ru.}

\begin{abstract}
We have measured the neutron electric dipole moment using spin rotation
in a non-centrosymmetric crystal. Our result is
$d_\mathrm{n} = (2.5\pm 6.5^\mathrm{stat}\pm 5.5^\mathrm{syst})\cdot 10^{-24}\,e\mathrm{cm}$.
The dominating contribution to the systematic uncertainty is
statistical in nature and will reduce with improved statistics.
The statistical sensitivity can be increased to
$2\cdot 10^{-26}\,e\mathrm{cm}$ in 100 days data taking with an
improved setup. We state technical requirements for a systematic
uncertainty at the same level.
\end{abstract}
\end{frontmatter}

\section{Introduction}
\label{Intr}
Electric dipole moments (EDMs) of elementary particles belong to the most
sensitive probes for CP violation beyond the Standard Model of Particle
Physics \cite{PospelovRitz2005}. Constraining or detecting EDMs of
different systems allows to gather experimental information about models
for new physics that is complementary to high energy physics data.

For the neutron EDM (nEDM), the most sensitive results
\cite{SussexEDM2006,pnpiedm} were obtained using ultracold neutrons
and Ramsey's resonance method. See \cite{LamoreauxGolub2009} for
a recent review of measurements using free neutrons.
Measurements using the interaction of neutrons with the atomic electric field\
in absorbing matter were pioneered by Shull and Nathans \cite{shull1967}. 
Abov and colleagues \cite{Abov1966} first discussed a spin-dependent term
in the scattering amplitude for neutrons in non-centrosymmetric non-absorptive
crystals. This term is caused by the interference of nuclear and spin-orbit
structure amplitudes. Forte \cite{forte1983} proposed to search for nEDM
related spin rotation in non-centrosymmetric crystals due to such
interference. In \cite{grav} it was pointed out that such
interference leads to a constant interplanar electric field affecting the
neutron during all time of its passage through the crystal. This field was
measured first in \cite{dfield}. The interference of nuclear and spin-orbit
structure amplitudes was tested by Forte and Zeyen \cite{ForteZeyen}
by spin rotation in a non-centrosymmetric crystal. 

Here we present the first measurement of the nEDM
based on an improved version of this method. Preliminary results and
a detailed description
of our method with references to earlier work have been published in 
conference proceedings \cite{dedm_test_NIM,dedm_test_PhysB}.

The statistical sensitivity of any experiment to measure the nEDM is
determined by the product $E\tau \sqrt{N}$, where $E$ is the value of
the electric field, $\tau$ the duration of the neutron's interaction
with the field and $N$ the number of counted neutrons. New projects
to measure the nEDM with UCNs aim to increase the UCN density and thus
$N$ by orders of magnitude and exploit in some cases the higher electric
field obtainable in liquid helium compared to vacuum
\cite{VanDerGrinten2009,Altarev2009}. In contrast, experiments with
non-centrosymmetric crystals exploit the interplanar electric field.
For quartz, this field was measured to be $E\approx 2\cdot10^8$\,V/cm
\cite{dfield}, several orders of magnitude higher than the electric
field achievable in vacuum or liquid helium. Furthermore, the
statistical sensitivity of the method profits from the higher flux of
the used cold neutrons, compared to UCNs available today. These
factors compensate the shorter interaction time of the neutrons with
the electric field, ultimately limited by the absorption in the crystal.
In \cite{LDM_sens} we have demonstrated that a statistical sensitivity
of $\sim 6\cdot 10^{-25}\,e\mathrm{cm}/\mathrm{day}$ can be
obtained, comparable to the most sensitive published UCN
experiments \cite{SussexEDM2006,pnpiedm}.

The results in \cite{LDM_sens} were obtained in Laue geometry that
is limited by systematics \cite{deformation}. Here we exploit the Bragg
geometry first proposed by Forte \cite{forte1983}. See
\cite{dedm_test_NIM} for a detailed comparison of the two schemes.

\section{Method}

We consider spin rotation for neutrons close to the Bragg condition
for the crystallographic plane $\bm{g}$ in a non-centrosymmetric
crystal. These neutrons are exposed
to an electric field $\bm{E} = \bm{E}_{\bm{g}}\cdot a$ where
$\bm{E}_{\bm{g}}$ is the interplanar electric field for plane $\bm{g}$
and $a$ describes the deviation of the neutron from the exact Bragg
condition (see \cite{dedm_test_NIM} for details). A nonzero nEDM
$d_\mathrm{n}$ results in neutron spin rotation by the angle
\begin{equation}
  \varphi_\mathrm{EDM} = \frac{2 E d_\mathrm{n} L}{
    \hbar v_\bot},
\label{Eq:fid}
\end{equation}
where $L$ is the length of the crystal and $v_\bot$ the component
of the neutron velocity perpendicular to the crystallographic plane
($L/v_\bot$ is the time the neutron interacts with the field).
By changing the deviation from the exact Bragg condition, $a$, value
and even sign of the effective electric field and the resulting spin
rotation $\varphi_\mathrm{EDM}$ can be selected.

On the other hand, the electric field causes a Schwinger magnetic field
in the rest frame of the neutron,
$\bm{H}_\mathrm{S}=[\bm{E}\times\bm{v}]/c$,
resulting in the spin rotation angle
\begin{equation}
\varphi_\mathrm{S} = \frac{2\mu_\mathrm{n} H_\mathrm{S} L}{\hbar v_ \bot}
   = \frac{2  E \mu_\mathrm{n}  L v_\parallel}{c\hbar v_\bot },\label{phis}
\end{equation}
where $\mu_\mathrm{n}$ is the neutron magnetic moment and
$v_\parallel$ the neutron velocity component parallel to the
crystallographic plane.

As $\bm{E}$ and $\bm{H}_\mathrm{S}$ are perpendicular to each other,
$\varphi_\mathrm{EDM}$ and $\varphi_\mathrm{S}$ can be separated by
three-dimensional (3D) polarisation analysis. Furthermore,
$\varphi_\mathrm{S}$ vanishes for Bragg angles of
$\pi/2$ ($v_\parallel=0$ in Eq.~(\ref{phis})).
This is used to suppress the effect due to Schwinger interaction:
The crystal is aligned such that the interplanar electric field
is parallel to the central neutron velocity, defining the $Z$
direction of a coordinate system. The incident neutron polarisation
is aligned in $X$ (or $Y$) direction. $\varphi_\mathrm{EDM}$ is
measured in the $X$-$Y$ plane. A residual Schwinger magnetic field
(for neutron trajectories deviating from the $Z$ direction or in
case of a slight misalignment of the crystal) has its largest
component in the $X$-$Y$ plane, thus creating a polarisation component
in $Z$ direction. Thus, $\varphi_\mathrm{S}$ can be derived
from the $Z$ component of the outgoing polarisation vector.
In summary, we measure $\varphi_\mathrm{EDM}$ from the component
$P_{XY}$ of the polarisation tensor and the residual Schwinger effect
from the components $P_{XZ}$ and $P_{YZ}$. $P_{YX}$, $P_{ZX}$ and
$P_{ZY}$ serve for control purposes.

In first order, the difference of the polarisation tensors for
positive and negative effective electric fields is:
\begin{equation}
{\bm{\Delta P}}  = g_{\rm n} \tau _0\left(
  \begin{array}{*{20}c}
   0 &
   -\left(H^z\frac{\Delta\tau}{\tau_0} + H_{\rm EDM}\right) &
   \left(H^y \frac{\Delta\tau}{\tau_0} + H_{\rm S}^y\right) \\
   \left(H^z \frac{\Delta\tau}{\tau_0} + H_{\rm EDM}\right) &
   0 &
   -\left(H^x\frac{\Delta\tau}{\tau_0} + H_{\rm S}^x\right) \\
   -\left(H^y\frac{\Delta\tau}{\tau_0} + H_{\rm S}^y\right) &
   \left(H^x \frac{\Delta\tau}{\tau_0} + H_{\rm S}^x\right) &
   0
  \end{array}
\right),
\end{equation}
{\noindent where $\tau_0=(\tau_+ + \tau_-)/2$, $\Delta\tau =(\tau_+ - \tau_-)/2$,
and $\tau_+$ and  $\tau_-$ are the times the neutrons stay in the
crystal for the positive and the negative electric field, respectively
(the neutron velocity is slightly different for $\pm a$). $H^i$ are
the components of the residual magnetic field and $H_\mathrm{S}^i$
the components of the Schwinger magnetic field $\bm{H}_\mathrm{S}$.
$g_n=2\mu_\mathrm{n}/\hbar=1.8\cdot 10^4 \mathrm{G}^{-1}\mathrm{s}^{-1}$
is the neutron gyromagnetic ratio,
$H_\mathrm{EDM}=E d_\mathrm{n}/\mu_\mathrm{n}$
the effective magnetic field corresponding to the electric field $E$.
For $E=1 \cdot 10^8$~V/cm and $d_n=10^{-26}\,e\mathrm{cm}$,
$H_\mathrm{EDM}=1.7\cdot 10^{-7}$\,G.}
  
\section{Experimental setup and procedure}\label{sec:setup}

\begin{figure}[bt] 
  \centering
  \includegraphics[width=0.9\columnwidth]{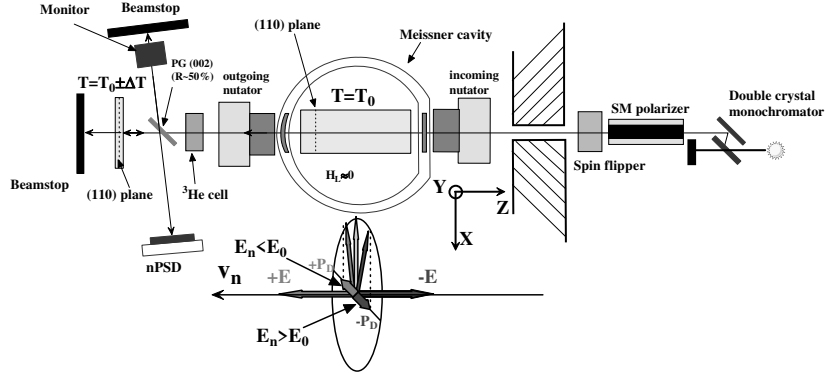}
  \caption{Scheme of the experiment. The neutrons come from the
    right. See Section~\ref{sec:setup} for details.}
  \label{setup} 
\end{figure}
The experiment was carried out at the cold neutron beam facility PF1B
\cite{Abele2006} of the Institut Laue-Langevin. A scheme of the setup
is shown in Fig.~\ref{setup}. Neutrons were
wavelength-preselected by a pyrolytic graphite monochromator
(adjusted to 4.91\,\AA, resolution about 1\%) and spin polarised 
to about 98\% by a super mirror polariser. A resonance spin flipper
permitted to flip the neutron polarisation. Cryopad \cite{Cryopad,CryopadEddy}
was used for 3D polarisation analysis. The direction of the incident
beam polarisation was oriented and the measured projection of the outgoing
polarisation vector selected by nutators (in the $X$-$Y$ plane)
and precession coils (inside the Meissner screen, for the $Z$ component).
The polarisation of the transmitted beam was analysed by a
$^3$He spin filter cell \cite{MeopILL2006} (measured polarisation value of the
unperturbed beam between 75\% and 87\%, depending on the $^3$He cell). The
currents in the precession coils were optimised experimentally
for the wavelength of the used neutrons.

We used the (110) reflection of a perfect quartz crystal of 14\,cm length. 
The effective angular mosaic spread of the crystal was
$\omega_\mathrm{m}\sim 1"$.
The neutrons with a well-defined deviation from the Bragg condition
for this crystal were selected behind the $^3$He spin filter cell
by back-reflection at a second quartz
crystal oriented parallel to the first one and at a slightly different
temperature. A shift of the neutrons by one Bragg width 
($\Delta\lambda_B/\lambda \approx 10^{-5}$ for the (110) plane of quartz)
corresponds to a temperature difference $\Delta T \approx \pm 1$\,K
(linear coefficient of thermal expansion for quartz
$\xi=\Delta L / L \approx 10^{ - 5}/\mathrm{K}$). Note that the absolute
temperature of the crystals is not important. The crystals were
temperature stabilised in individual housings connected by a
common water circuit. The temperature of the second crystal was
varied by Peltier elements.

The back-reflected neutrons were separated from the beam by a pyrolytic
graphite monochromator (PG (002)) with a reflectivity of about 50\% and
directed to a position sensitive detector (nPSD). The nPSD allowed us to
analyse data depending on the deviation of the neutron's path from
the $Z$ direction: Larger deviations correspond to larger $v_\parallel$ and
thus $\varphi_\mathrm{S}$. This monochromator unavoidably reflects about 50\%
of the neutrons after the spin filter, before the second quartz crystal.
These neutrons were used to monitor intensity and polarisation.

During installation, the crystals were pre-aligned optically using an
autocollimator. Final alignment was performed with neutrons,
allowing for the possibility of slightly different lattice
constants of the crystals at the same temperature: The
intensity of the neutron beam backscattered by the second crystal
has a minimum if the Bragg condition of the second crystal (varied by
the temperature difference) corresponds to the Bragg condition of
the first one. If the crystal planes of both crystals are parallel,
this minimum has the smallest width. The alignment with the minimum
width was found by scanning the temperature difference for different
angles (in $X$ and $Y$ direction) of the second crystal. Data
show that the lattice constants were equal for $\Delta T=-0.8$\,K
\cite{dedm_test_NIM},
which can be attributed to slightly different impurities of the
two crystals. From previous measurements \cite{Bragg_test} we selected
the working points with maximal electric field magnitude
$\Delta T_-=-2.0$\,K and $\Delta T_+=+0.4$\,K.

The different measured combinations of electric field direction and
components of the polarisation tensor are $P_{ij\pm}$, where
the index $\pm$ refers to the temperature difference
of the crystals $\Delta T_\pm$. The indices $i$ and $j$
indicate the directions of the incident and outgoing neutron
polarisation selected by Cryopad.
Each component $P_{ij\pm}$ was measured for 40\,s by flipping
the incident neutron polarisation every 1-4\,s with
the resonance flipper (states $\uparrow$ off, $\downarrow$ on),
in a drift-compensating scheme $\uparrow
\downarrow\downarrow\uparrow\downarrow\uparrow\uparrow\downarrow$.
The different states $P_{ij\pm}$ were measured in drift-compensating
cycles, one for the EDM effect, $P_{XY-}$, $P_{XY+}$, $P_{XY+}$,
$P_{XY-}$, $P_{YX-}$, $P_{YX+}$, $P_{YX+}$ ,$P_{YX-}$, and one for the
Schwinger effect, $P_{XZ-}$, $P_{YZ-}$, $P_{XZ+}$, $P_{YZ+}$, $P_{YZ+}$,
$P_{XZ+}$, $P_{YZ-}$, $P_{XZ-}$. EDM data were taken 2/3 of the total
time, Schwinger data due to the smaller required statistics only 1/3.
The dominating drift originated from the depolarisation of the $^3$He
cells with time (time constants large compared to a single cycle).

\section{Results}

\begin{figure}[bt]
  \centering
  \includegraphics[width=0.6\columnwidth]{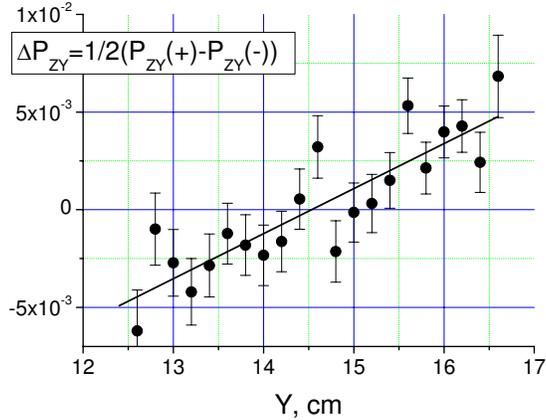}
  \caption{Measured dependence of $\Delta P_{ZY}$ on the spatial coordinate
    $Y$ of the nPSD.}
    \label{fig:DPZY}
\end{figure}

From the measured data we determined the elements of the difference of the
polarisation matrices,
$\Delta P_{ij}=1/2 (P_{ij+}-P_{ij-})$. The Schwinger data 
showed the expected linear variation over the nPSD: neglecting
the spatial extension of the beam, the deviation of a neutron
from the centre of the nPSD is proportional to $v_\parallel$.
Fig.~\ref{fig:DPZY} shows a linear fit to the data $\Delta P_{ZY}$.
This allows us to derive the electric field acting on the
neutrons:
\begin{equation}
  E_\mathrm{exp}=(0.7\pm 0.1)\cdot 10^{8}\,\mathrm{V/cm}.
\end{equation}
This value is consistent with our preliminary result obtained for 
the Bragg angle $\theta_B=86^\circ$ \cite{Bragg_test}.

\begin{figure}[bt]
  \centering
    \includegraphics[width=0.6\columnwidth]{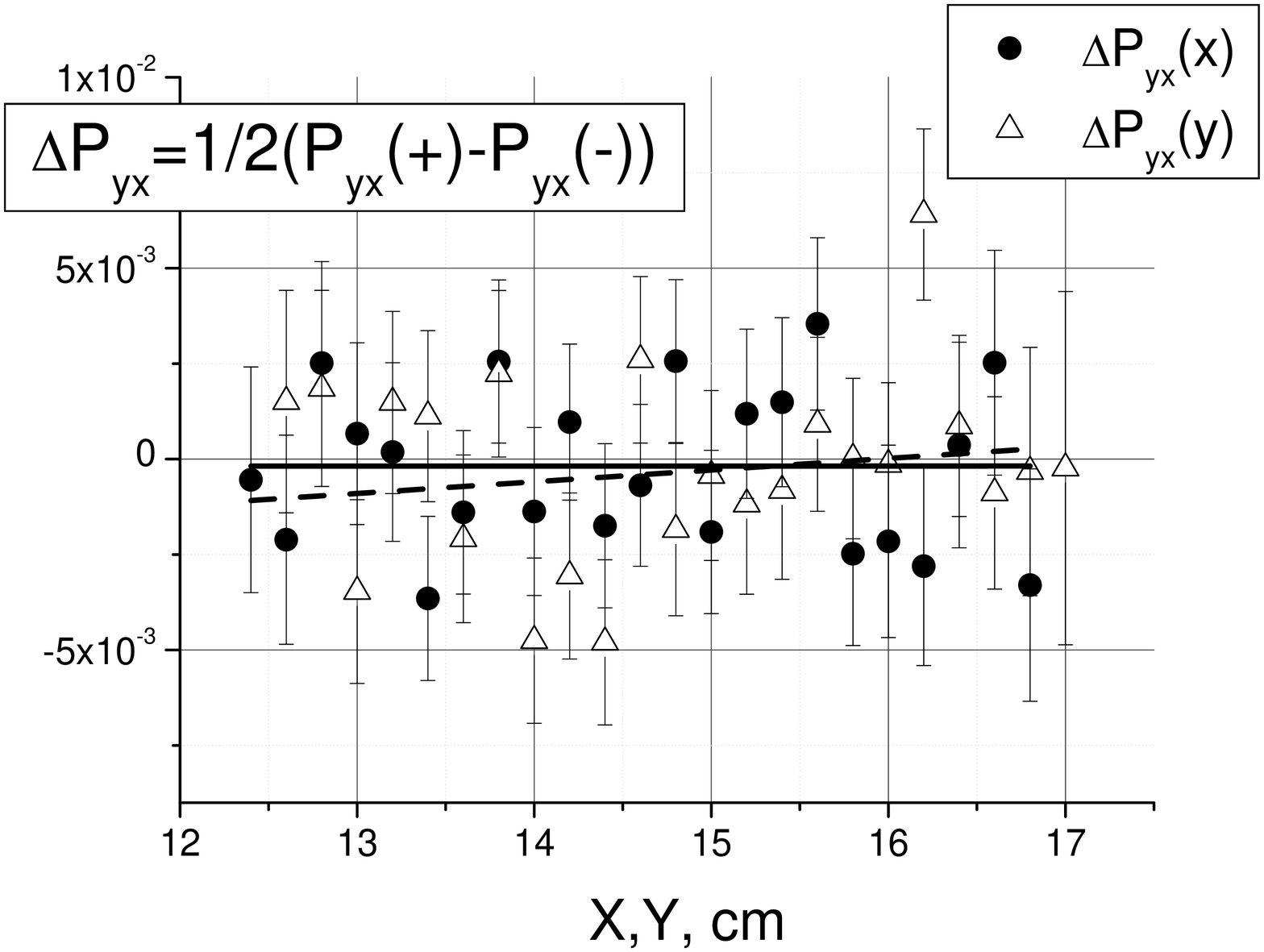}
    \includegraphics[width=0.6\columnwidth]{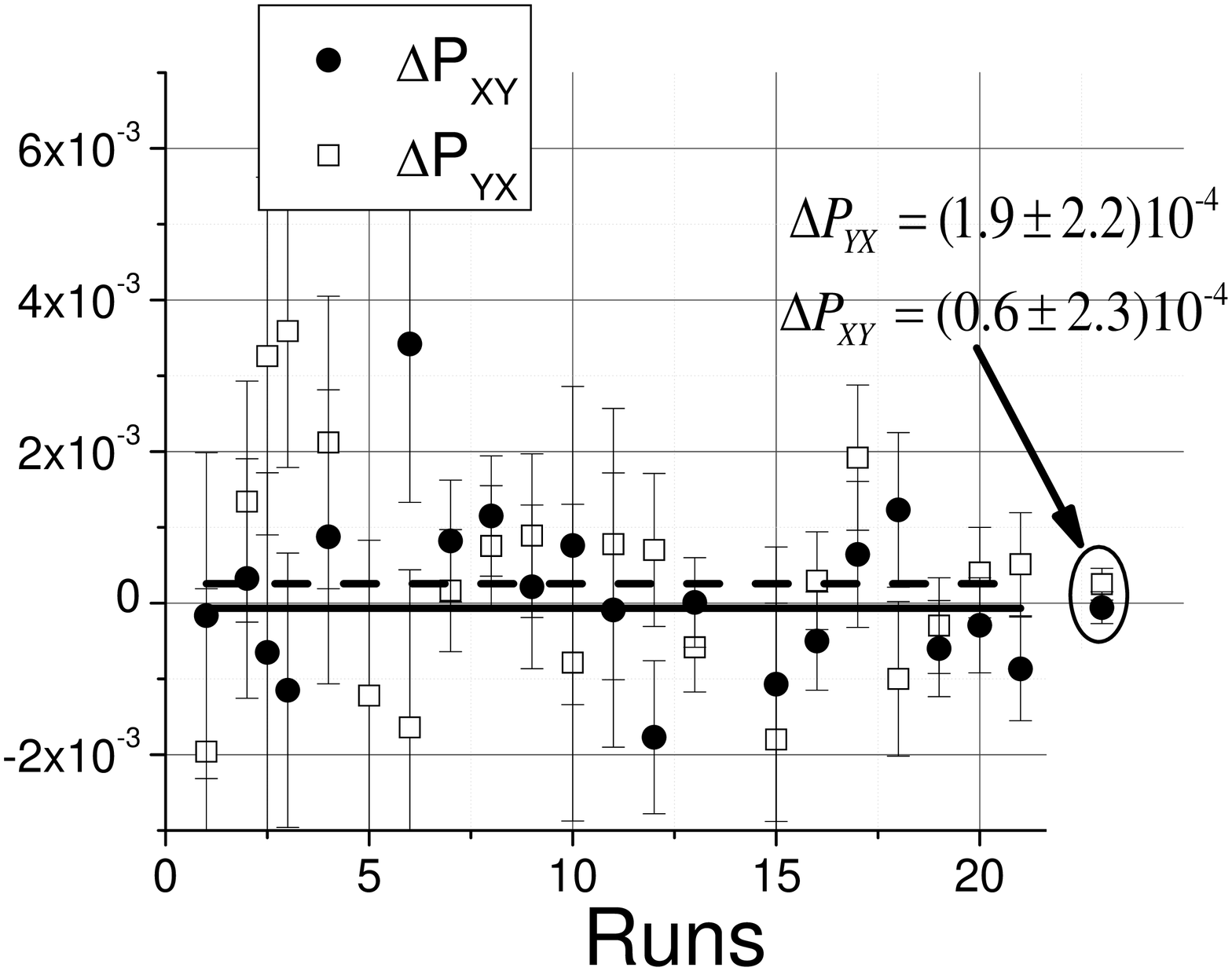}
  \caption{Top: Measured dependences of $\Delta P_{YX}$ on the spatial
    coordinates $X$ and $Y$ of the nPSD. Bottom: Results of the different
    measurements of $\Delta P_{XY}$ and $\Delta P_{YX}$, used to derive
    the nEDM.}
    \label{fig:PyxnEDM}
\end{figure}

Fig.~\ref{fig:PyxnEDM} (top) shows an example for the spatial dependence
of the matrix element $\Delta P_{YX}$ sensitive to the nEDM.
The summary of all accumulated data for the nEDM is
presented in Fig.~\ref{fig:PyxnEDM} (bottom). The average value
for the nEDM spin rotation is
\begin{equation}
  \varphi_\mathrm{EDM}  = (0.9 \pm 2.3)\cdot 10^{-4}\,\mathrm{rad}, 
\end{equation}
corresponding to the nEDM (see Eq.~(\ref{Eq:fid}))
$d_\mathrm{n} = (2.5\pm 6.5)\cdot 10^{-24}\,e\mathrm{cm}$.

The slope of the spatial dependencies of $\Delta P_{YX}$ and
$\Delta P_{XY}$ is related to the Schwinger effect. It was found
to be consistent with zero and used to estimate a related
systematic uncertainty of
$\sigma d_\mathrm{n,S} < 5\cdot 10^{-24}\,e\mathrm{cm}$.
Note that this uncertainty is smaller than the statistical uncertainty
of the nEDM taken from the same data and would reduce with better
statistics. It includes the error caused by
imperfect angular accuracy of the 3D polarisation
analysis (if the analysed $Z$ component is not perpendicular to the
$X$-$Y$ plane, the corresponding projection of the Schwinger effect
would contribute to $\Delta P_{XY}$ and $\Delta P_{YX}$).

The residual magnetic field inside Cryopad was $H\sim (1-2)$\,mG.
The reversal of the electric field leads to a small change of the
neutron wavelength and thus to a difference of the time the neutron
stays inside Cryopad. This time difference was measured directly:
$\Delta \tau /\tau < 5 \cdot 10^{-4}$. The combination of the
residual magnetic field and the time difference results in a
systematic uncertainty of
$\sigma d_{\mathrm{n},H} < 4.5\cdot 10^{-25}\,e\mathrm{cm}$.  

The time stability of the experiment was controlled by the beam
monitor, which registered neutrons behind Cryopad without
interaction with the second quartz crystal.
The average electric field acting on these neutrons was about zero
(taking into account the wavelength distribution in the neutron beam,
it was $\sim 10^3$ times less than the field for the neutrons
that would be selected by the second quartz crystal). The difference
in polarisation measured by the beam monitor between the two
working points, $\Delta P < 2\cdot 10^{-5}$, can be used to limit
a possible systematic effect due to beam fluctuations in time,
$\sigma d_\mathrm{n,time} <10^{-24}\,e\mathrm{cm}$. 

\begin{figure}[bt]
  \centering
  \includegraphics[width=0.6\columnwidth]{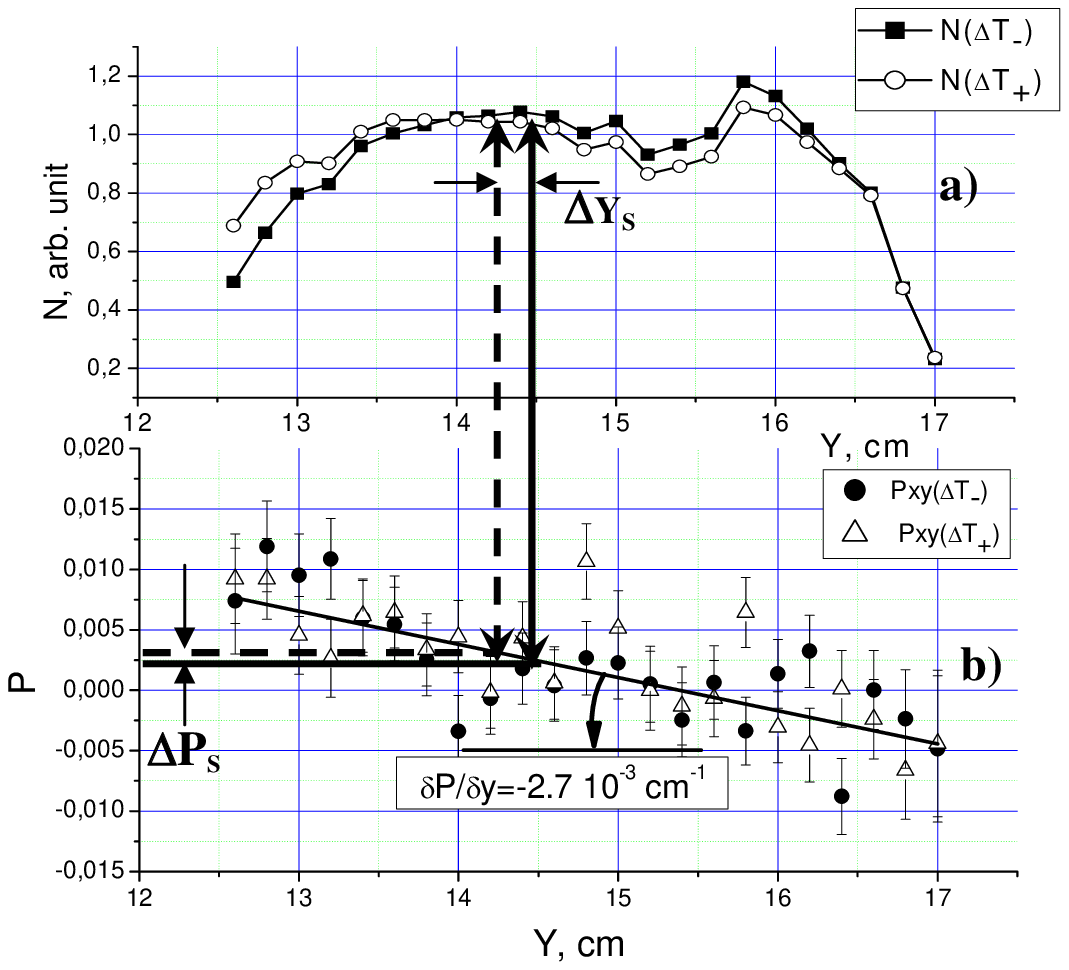}
  \caption{Spatial distribution of the registered neutron
    intensity a) and measured polarisation b) for the two working
    points $\Delta T_\pm$.}
  \label{fig:17}
\end{figure}
We have observed a spatial shift of the neutron beam at the
nPSD correlated with the working point $\Delta T_\pm$ (see
Fig.~\ref{fig:17}a), caused by an imperfect alignment
of the two quartz crystals, and a spatial dependence of the
polarisation (see Fig.~\ref{fig:17}b), probably due to the cylindrical
shape of the outgoing window of Cryopad and the presence of a
residual magnetic field. The combination of both effects
causes a systematic error of
$\Delta P_\mathrm{s}=1/2\,\Delta Y_\mathrm{s}\,
\delta P/\delta Y_\mathrm{s}$.
For the final crystal alignment $\Delta Y_\mathrm{s} = 0.03$\,cm this
corresponds to $\Delta P_\mathrm{s}=4 \cdot 10^{-5}$ and
$\sigma d_\mathrm{n,s} = 2\cdot 10^{-24}\,e\mathrm{cm}$. 

Spin rotation caused by the weak interaction cancels for the two crystal
temperatures in first order. The cancellation is incomplete because of
a cross term of weak and Schwinger rotation, but the uncompensated
part is below $10^{-4}$ of the weak spin rotation and can be neglected.

Summing these independent systematic errors we obtain the final result
\begin{equation}
  d_\mathrm{n} = (2.5\pm6.5^\mathrm{stat}\pm5.5^\mathrm{syst})\cdot
    10^{-24}\,e\mathrm{cm}.
\end{equation}
Note that the systematic error is mainly statistical in nature.
This result is about two orders of magnitude less precise than
present experiments using UCNs and Ramsey's resonance method.

\section{Outlook}

\emergencystretch20pt

The statistical sensitivity of the experiment described above
was $1.6\cdot 10^{-23}\,e\mathrm{cm/day}$. In \cite{dedm_test_NIM}
we have shown that the sensitivity can be improved by a factor 65
to about $2.5\cdot 10^{-25}\,e\mathrm{cm/day}$, essentially by increasing
the divergence acceptance of the installation, the beam size,
and the crystal length. This improved sensitivity is comparable to
state-of-the-art nEDM measurements using UCNs and permits
a statistical uncertainty at the $10^{-26}\,e\mathrm{cm}$
level in a beam time of 100 days. A further improvement may be
possible by using crystals with higher interplanar electric
fields. However, these crystals are presently not available in
the required quality and size.

In Table~\ref{Tab:Req} we list the experimental parameters that are
required for an adequate systematic uncertainty. The values are
achievable. Note that a
variation of the interplanar distance in the crystal of the order
of the Bragg width or larger does not simulate an nEDM but reduces
the magnitude of the effective electric field.
\begin{table}[bt]
  \centering
  \caption{Experimental parameters for a systematic uncertainty of
    $\sigma d_\mathrm{n} \sim 10^{-26}\,e\mathrm{cm}$.}
  \label{Tab:Req}
  \begin{tabular}{|l|l|l|}
  \hline
  \multicolumn{2}{|l|}{\textbf{Parameter}} & \textbf{Value} \\ \hline
  \multicolumn{2}{|l|}{Interplanar distance variation $\Delta d/d$}  &
    $< 10^{-5}$\\ \hline
  \multicolumn{2}{|l|}{Residual magnetic field} &  \\ \hline
  ~~~~~ & Value & $ \sim 10^{ - 4}$\,G \\ \cline{2-3} 
  ~~~~~ & Time stability & $ \sim 10^{-5}$\,G/h \\ \hline
  \multicolumn{2}{|l|}{Accuracy of 3D polarisation analysis}  &
    $\sim 10^{-3}$\,rad\\ \hline
  \multicolumn{2}{|l|}{Angular crystals alignment}  & $\sim 0.01^\circ$
    \\ \hline
  \multicolumn{2}{|l|}{Precision of temperature control} & $\sim 0.01$\,K
    \\ \hline
  \multicolumn{2}{|l|}{Flatness of Cryopad windows} &
    $\sim 10^{-4}$\,rad\\ \hline
\end{tabular}
\end{table}

We conclude that neutron spin rotation in quartz
in Bragg geometry permits to measure the nEDM with a precision
comparable to existing UCN experiments but using a completely
different method. Efforts are now being made to produce larger crystals
and a version of Cryopad optimised for that kind of measurements.

\paragraph*{Acknowledgements}
The authors would like to thank the personnel of the ILL reactor
(Grenoble, France) for technical assistance in the experiment.
This work was supported by RFBR (grants No 09-02-00446).

\end{document}